\documentclass[conference,10pt]{IEEEtran}
\IEEEoverridecommandlockouts
\usepackage{graphicx}
\usepackage{caption}
\usepackage{enumerate}
\usepackage{epsfig}
\usepackage{epstopdf}
\usepackage{svg}
\usepackage{cite}
\usepackage{subfigure}
\usepackage{amsmath,amssymb,amsfonts}
\usepackage{algorithmic}
\usepackage{tabularx}
\usepackage{textcomp}
\usepackage{xcolor}
\usepackage{stfloats}
\usepackage{multirow}
\usepackage{url}
\usepackage[ruled]{algorithm2e}
\usepackage{etoolbox,xstring,mfirstuc,textcase}
\usepackage{bm}
\usepackage{amsthm}
\usepackage{stackengine}
\usepackage{cuted}
\usepackage{makecell}
\usepackage{booktabs}
\usepackage{threeparttable}
\usepackage[skip=3pt]{caption}
\usepackage[explicit]{titlesec}

\begin{document}

\title{Baseband-Free Wideband MU-MIMO OFDM via Stacked Intelligent Metasurfaces}

\author{Zheao Li, Jiancheng An, and Chau Yuen\textsuperscript{*} \\
School of Electrical and Electronic Engineering, Nanyang Technological University, 639798, Singapore.\\
		Email: zheao001@e.ntu.edu.sg,~jiancheng\_an@163.com,~chau.yuen@ntu.edu.sg.
\vspace{-0.6 cm}
}
\maketitle
\begin{abstract}
This paper proposes a novel baseband-free wideband MU-MIMO OFDM transmitter architecture enabled by two cascaded stacked intelligent metasurfaces (SIMs). Unlike conventional wireless transmitters, the proposed design shifts symbol loading, MU-MIMO precoding, and OFDM modulation from digital baseband to the wave domain. Specifically, in the first SIM (SIM$_1$), a programmable symbol-loading interface first loads one stacked virtual-subcarrier coefficient vector onto a common monochromatic carrier, while the remaining layers perform channel-adaptive MU-MIMO precoding from the data streams to the port-subcarrier domain. Then, the second SIM (SIM$_2$) realizes offline-configured wave-domain OFDM modulation by implementing the inverse discrete Fourier transform (IDFT) and cyclic-prefix (CP) insertion operator directly in the wave domain. Based on this architecture, we develop a wave-domain system model that explicitly characterizes the virtual-to-physical transition from block-level coefficients to radiated OFDM subcarriers. The designs of SIM$_1$ and SIM$_2$ are formulated as two operator-fitting problems, respectively targeting a block-diagonal wideband precoder and the ideal OFDM modulation operator. Under practical discrete phase constraints, we develop a quantization-aware training-based gradient descent (QAT-GD) framework for both SIM stages. Numerical results verify the effectiveness of the proposed optimization, the correctness of the synthesized wave-domain functionalities, and the strong end-to-end MU-MIMO OFDM performance of the resulting architecture.
\end{abstract}

\begin{IEEEkeywords}
Baseband-free communications, stacked intelligent metasurfaces (SIM), wave-domain computing, fully-analog precoding, MU-MIMO, OFDM.
\end{IEEEkeywords}
\vspace{-0.3 cm}
\section{Introduction}

Modern high-speed wireless communication systems are increasingly expected to support massive data traffic, high spectral efficiency, low latency, and reliable multiuser connectivity. To meet these requirements, multiple-input multiple-output (MIMO) transmission and multicarrier modulation have become two foundational technologies in contemporary wireless networks. In particular, orthogonal frequency-division multiplexing (OFDM) is widely adopted to combat frequency selectivity in wideband channels, while multiuser MIMO (MU-MIMO) enables spatial multiplexing and interference management across multiple users  \cite{OFDM}. As a result, MU-MIMO OFDM has emerged as a standard physical-layer architecture for modern broadband wireless systems. However, it strongly relies on two computationally intensive digital baseband modules: subcarrier-wise MU-MIMO precoding in the spatial domain and OFDM waveform synthesis via inverse discrete Fourier transform (IDFT) and cyclic-prefix (CP) insertion in the time domain. As the antenna, stream, and subcarrier dimensions grow, these operations incur substantial hardware cost, computing complexity, and power consumption, especially when high-resolution digital-to-analog converters (DACs) and multiple radio-frequency (RF) chains are required for large-scale wideband transmission.

Recently, stacked intelligent metasurfaces (SIMs) have emerged as a promising platform for electromagnetic (EM) wave signal processing and computing \cite{SIM0, SIM}. Owing to successive propagation and programmable phase across stacked layers, SIMs can emulate high-dimensional matrix transformations directly in the EM wave domain and have shown strong potential for beamforming in physical layer security and cell-free networks \cite{SIM1, SIM2, SIM3, SIM4}. In wireless communications, existing SIM-enabled designs have demonstrated encouraging gains for narrowband multiuser beamforming and near-field spatial processing \cite{SIMsur}, suggesting that SIM can serve as a low-power analog computing medium for high-dimensional propagation environment shaping. SIM can also be employed to support many beyond-communication functionalities, such as sensing, localization, and image recognition \cite{SIM5,SIMsur2}.

However, current SIM-based communication architectures still fall short of a fully wave-domain wideband transmitter. Most existing studies remain essentially narrowband, where the metasurfaces are only required to emulate a spatial-domain precoding operator. A few recent wideband works have shown that SIM can realize wideband MIMO communications \cite{Wideband-SIM1, Wideband-SIM2, Wideband-SIM3}, but the OFDM modulator itself is still generated in digital baseband. This limitation is fundamental: unlike narrowband beamforming, OFDM modulation is a more structured and computationally involved operation, since it must jointly transform a set of subcarrier coefficients into a CP-extended time-domain waveform. Consequently, a truly fully wave-domain wideband architecture that simultaneously realizes MIMO precoding and OFDM modulation remains largely unexplored.

In this paper, we address this gap by proposing a novel baseband-free wideband MU-MIMO OFDM transmitter architecture composed of two cascaded SIMs, denoted by SIM$_1$ and SIM$_2$. Rather than treating SIMs as add-on modules to an otherwise conventional communication chain, the proposed design redistributes the key transmitter-side baseband functions in the wave domain with different adaptation timescales. We develop an operator-level system model based on a virtual-subcarrier representation, where SIM$_1$ realizes programmable symbol loading and block-diagonal wideband precoding, while SIM$_2$ realizes an IDFT and CP insertion operator, thereby explicitly characterizing the transition from virtual port-subcarrier coefficients to physically radiated OFDM waveforms. Under practical discrete phase constraints, we develop a quantization-aware training-based gradient descent (QAT-GD) framework for configuring both SIMs and validate the proposed architecture from three complementary perspectives: optimization effectiveness, functionality realization, and communication performance. In other words, this work introduces a new transmitter realization principle, where the conventional transmitter-side baseband chain is redistributed across programmable wave-domain stages.

\section{System Model}

In this section, we consider a SIM-enhanced wideband MU-MIMO OFDM downlink transmitter composed of SIM$_1$ and SIM$_2$, both illuminated by a common monochromatic carrier at frequency $f_c$. The key idea is to move the high-dimensional operations of a conventional wideband transmitter, including MU-MIMO precoding and OFDM waveform synthesis, from digital baseband to programmable wave-domain propagation.

\subsection{SIM-enhanced Wave-Domain MU-MIMO OFDM System}
\label{subsec:system_overview}

\begin{figure*}
	\centerline{\includegraphics[width=1\textwidth]{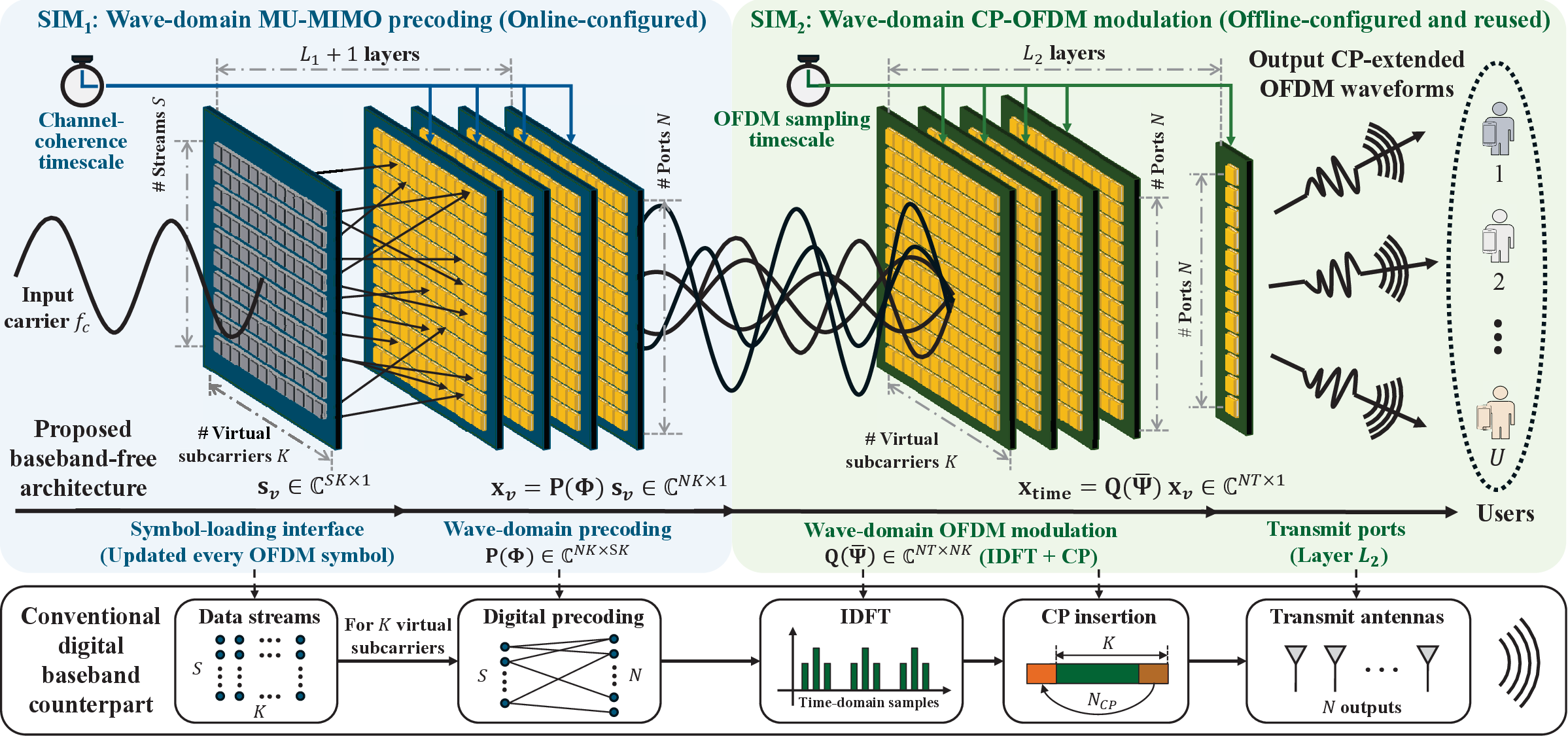}}
	\caption{Proposed baseband-free wideband MU-MIMO OFDM transmitter.}
	\label{Fig0}
\vspace{-0.3 cm}
\end{figure*}

Unlike a conventional wideband architecture where symbol loading, MU-MIMO precoding, and OFDM modulation are carried out digitally before radiation, the proposed design in Fig.~\ref{Fig0} redistributes all three transmitter-side baseband functions to the EM wave domain enhanced by two cascaded SIMs under a common monochromatic carrier excitation at frequency $f_c$. A key architectural feature is that the three wave-domain processing stages operate at different timescales. Layer 0 of the SIM$_1$ is updated at the OFDM-symbol level to load the current coefficient vector, while the rest of the layers are updated at the channel-coherence-block timescale so that their phase configuration can track the current channel for MU-MIMO precoding. By contrast, SIM$_2$'s phase schedule is designed offline to approximate the IDFT-and-CP operator, and the same schedule is then reused during transmission. Therefore, the proposed architecture separates data adaptation, channel adaptation, and OFDM modulation across three distinct wave-domain stages.

SIM$_1$ consists of one programmable loading layer indexed by $0$, followed by $L_1$ transmissive metasurface layers indexed by $\mathcal{L}_1=\{1,\ldots,L_1\}$. Layer 0 loads one stacked virtual-subcarrier coefficient vector, yielding the input vector $\mathbf{s}_{{v}}\in\mathbb{C}^{SK\times1}$, where $S$ is the number of streams and $K$ is the number of subcarriers. The subsequent layers 1 to $L_1$ collectively realize the fully-analog MU-MIMO precoding operator $\mathbf{P}(\mathbf{\Phi})$, which maps the virtual stream block to the virtual port-subcarrier coefficients $\mathbf{x}_{\mathrm{v}}\in\mathbb{C}^{NK\times 1}$:
\begin{equation}
\mathbf{x}_{{v}}
=
\mathbf{P}(\mathbf{\Phi}){\mathbf{s}}_{{v}}
=
\big[\mathbf{x}^{\top}[0],\mathbf{x}^{\top}[1],\ldots,\mathbf{x}^{\top}[K-1]\big]^T,
\label{eq:xv_partition_new}
\end{equation}
where $\mathbf{x}[k]=\mathbf{P}[k]\mathbf{s}[k]\in\mathbb{C}^{N\times1}$ is the virtual port-domain coefficient vector associated with the $k$-th virtual subcarrier, and $N$ is the number of the transmit port after precoding.

SIM$_2$ consists of $L_2$ transmissive metasurface layers, indexed by $\mathcal{L}_2=\{1,\ldots,L_2\}$. Its role is to convert the port-subcarrier vector $\mathbf{x}_{ v}$ into the actual CP-extended OFDM waveform, where the final layer acts as transmit ports. Different from SIM$_1$, SIM$_2$ is not adapted to the instantaneous channel or the data block. Instead, its phase schedule is designed offline to fit the target IDFT-and-CP operator and is then fixed during online transmission. Hence, SIM$_2$ acts as a static waveform modulation module from the system perspective, although its operation internally follows a prescribed sequence across the $T$ samples of each OFDM block. This stage realizes the wave-domain OFDM modulation operator $\mathbf{Q}(\bar{\mathbf{\Psi}})$, which transforms the virtual port-subcarrier coefficients into the radiated CP-extended OFDM waveform $\mathbf{x}_{\mathrm{time}}\in\mathbb{C}^{NT\times1}$:
\begin{align}
    \mathbf{x}_{\mathrm{time}}
    =
    \mathbf{Q}(\bar{\mathbf{\Psi}})\,\mathbf{x}_{{v}}
=
    \begin{bmatrix}
        \mathbf{x}_{\mathrm t}^\top[0],
        \mathbf{x}_{\mathrm t}^\top[1],
        \cdots,
        \mathbf{x}_{\mathrm t}^\top[T-1]
    \end{bmatrix}^\top,
    \label{eq:x_time_Q_tot}
\end{align}
where $\mathbf{x}_t[t]\in\mathbb{C}^{N\times1}$ denotes the $N$-dimensional port-domain output at time slot $t$, and the $T$ stacked outputs jointly form one radiated CP-extended OFDM waveform.

By designing two SIMs to realize the desired responses in the wave domain, the actual multicarrier waveform is generated after SIM$_2$. After CP removal and DFT operation at the receivers, the same index $k$ is then interpreted as the physical subcarrier index. This virtual-to-physical transition connects the wave-domain architecture to the conventional wideband OFDM channel model. More detailed architectures are expressed in the following two subsections.
\vspace{+0.05 cm}
\subsubsection{SIM$_1$-Based Symbol Loading and MU-MIMO Precoding}
\vspace{+0.05 cm}
According to \cite{MultiMod} and \cite{MultiMod2}, the programmable metasurface layer 0 of SIM$_1$ can load the coefficient block $\{{\mathbf{s}}[k]\}_{k\in\mathcal{K}}$ onto the common carrier $f_c$ , and inject the resulting wave-domain excitation into the subsequent cascades. For one stacked virtual-subcarrier coefficient vector, let $\mathbf{s}[k]\in\mathbb{C}^{S\times1}$ denote the $S$-stream coefficient vector associated with virtual-subcarrier index $k\in\mathcal{K}=\{0,1,\ldots,K-1\}$. Stacking all virtual-subcarrier coefficients yields
\begin{equation}
\mathbf{s}_{{v}}
=
\big[
\mathbf{s}^{\top}[0],\mathbf{s}^{\top}[1],\ldots,\mathbf{s}^{\top}[K-1]
\big]^\top
\in\mathbb{C}^{SK\times1}.
\end{equation}

Let $\mathbf{W}^1\in\mathbb{C}^{NK\times SK}$ denote SIM$_1$'s effective loading-and-coupling matrix from the $SK$ loading ports generated by the 0-th layer to the 1-th layer. This matrix abstracts the combined effect of symbol loading, carrier-domain modulation, and near-field coupling from layer $0$ to layer $1$. For $l=2,\ldots,L_1$, let $\mathbf{W}^{l}\in\mathbb{C}^{NK\times NK}$ denote the inter-layer propagation matrix between the $(l\!-\!1)$-th and $l$-th SIM$_1$ layers. Since SIM$_1$ is monochromatic, all internal propagation matrices are evaluated at the same carrier frequency $f_c$. Their entries of $\mathbf{W}^{l}$ follow the Rayleigh--Sommerfeld form \cite{RS}
\begin{equation}
    [\mathbf{W}^{l}]_{m,m'}
    =
    \frac{A_1 d_1}{(r_{m,m'}^{l})^2}
    \left(
        \frac{1}{2\pi r_{m,m'}^{l}}
        -j\frac{f_c}{c}
    \right)
    e^{j2\pi r_{m,m'}^{l}f_c/c},
    \label{eq:W_new}
\end{equation}
where $m,m'\in\mathcal{M}=\{1,\ldots,NK\}$ index the corresponding meta-atoms or output ports, $A_1$ is the meta-atom area, $d_1$ is the inter-plane spacing, $r^{l}_{m,m'}$ is the corresponding propagation distance, and $c$ is the speed of light.

The diagonal phase shift matrix $\mathbf{\Phi}^{l} \in\mathbb{C}^{NK\times NK}$ of the $l$-th SIM$_1$ layer is modeled as
\begin{equation}
    \mathbf{\Phi}^{l}
    =
    \operatorname{diag}
    \big(
        e^{j\theta_1^{l}},\ldots,e^{j\theta_{NK}^{l}}
    \big),
    \quad \forall\ell\in\mathcal{L}_1,
    \label{eq:Phi_new}
\end{equation}
where $\theta_m^{l}\in[0,2\pi)$ is the phase shift applied by the $m$-th meta-atom on layer $l$.

Accordingly, the end-to-end SIM$_1$ operator is written as
\begin{equation}
    \mathbf{P}(\mathbf{\Phi})
    =
    \mathbf{\Phi}^{L_1}
    \mathbf{W}^{L_1}
    \cdots
    \mathbf{\Phi}^{1}
    \mathbf{W}^{1}
    \in \mathbb{C}^{NK\times SK}.
    \label{eq:P_new}
\end{equation}

To preserve the conventional OFDM coefficient structure before waveform modulation, SIM$_1$ is intentionally designed to realize only stream-to-port mapping for each subcarrier index. Accordingly, the intended SIM$_1$ operator takes the block-diagonal form
\begin{equation}
\mathbf{P}(\mathbf{\Phi})
=
\operatorname{blkdiag}\!\big(
\mathbf{P}[0],\mathbf{P}[1],\ldots,\mathbf{P}[K-1]
\big),
\label{eq:pp}
\end{equation}
where $\mathbf{P}[k]\in\mathbb{C}^{N\times S}$ denotes the effective precoding block associated with virtual subcarrier index $k$.
\vspace{+0.05 cm}
\subsubsection{SIM$_2$-Based Wave-Domain OFDM Modulation}
\vspace{+0.05 cm}
$\mathbf{U}^{1}\in\mathbb{C}^{NK\times NK}$ denotes the propagation matrix from the final layer of SIM$_1$ to the first input ports of SIM$_2$. For $\ell=2,\ldots,L_2-1$, let $\mathbf{U}^{\ell}\in\mathbb{C}^{NK\times NK}$ denote the inter-layer propagation matrix of the $(\ell\!-\!1)$-th and $\ell$-th SIM$_2$ layers, and and $\mathbf{U}^{L_2}\in\mathbb{C}^{N\times NK}$ maps the $(L_2\!-\!1)$-th layer to the $L_2$-th $N$-port output layer of SIM$_2$. Again, all propagation matrices $\mathbf{U}^{\ell}$ are with entries
\begin{equation}
    [\mathbf{U}^{\ell}]_{\mu,\mu'}
    =
    \frac{A_2 d_2}{(r_{\mu,\mu'}^{\ell})^2}
    \left(
        \frac{1}{2\pi r_{\mu,\mu'}^{\ell}}
        -j\frac{f_c}{c}
    \right)
    e^{j2\pi r_{\mu,\mu'}^{\ell}f_c/c},
    \label{eq:U_entry}
\end{equation}
where $\mu,\mu'\in\mathcal{M}$, $A_2$, $d_2$, and $r_{\mu,\mu'}^{\ell}$ are defined analogously for SIM$_2$.

Over one CP-extended OFDM block of length $T = K + N_{\mathrm{CP}}$, where $N_{\mathrm{CP}}$ is the CP length, SIM$_2$ switches across the $T$ discrete time indices $t\in\mathcal{T}\triangleq\{0,1,\ldots,T-1\}$. The phase shift matrix $\mathbf{\Psi}^{\ell}[t] \in\mathbb{C}^{NK\times NK}$ of the $\ell$-th SIM$_2$ layer at $t$ is
\begin{equation}
    \mathbf{\Psi}^{\ell}[t]
    =
    \operatorname{diag}
    \big(
        e^{j\zeta_1^{\ell}[t]},\ldots,e^{j\zeta_{NK}^{\ell}[t]}
    \big),
   \quad \ell=1,\! \cdots,\!{L}_2\!-\!1,
    \label{eq:Psi}
\end{equation}
where $\zeta_\mu^{\ell}[t]\in[0,2\pi)$ denotes the phase shift applied by the $\mu$-th meta-atom on layer $\ell$. $\mathbf{\Psi}^{L2}[t]\in\mathbb{C}^{N\times N}$ is the phase shift matrix of the $L_2$-th layer as transmit ports.

The instantaneous operator at time index $t$ is
\begin{equation}
    \mathbf{Q}[t]=\mathbf{Q}(\mathbf{\Psi}[t])
    =
    \mathbf{\Psi}^{L_2}[t]
    \mathbf{U}^{L_2}
    \cdots
    \mathbf{\Psi}^{1}[t]
    \mathbf{U}^{1}
    \in\mathbb{C}^{N\times NK},
    \label{eq:Qinst_bp}
\end{equation}
where $\mathbf{Q}[t]$ naturally represents the wave-domain OFDM modulation process from the $NK$-dimensional input ports to the $N$ physical transmit ports at each time slot $t$.

Collecting all time-varying phase matrices yields $
    \bar{\mathbf{\Psi}}
    \triangleq
    \left\{
        \mathbf{\Psi}^{\ell}[t]
    \right\}_{\ell\in\mathcal{L}_2,\,t\in\mathcal{T}}$. Stacking the operators over all $T$ time indices gives the overall time-domain modulation operator
\begin{equation}
    \mathbf{Q}(\bar{\mathbf{\Psi}})
    \triangleq
    \begin{bmatrix}
        \mathbf{Q}^\top[0],
        \cdots,
        \mathbf{Q}^\top[T-1]
    \end{bmatrix}^\top
    \in\mathbb{C}^{NT\times NK}.
    \label{eq:Qall_SIM2_B}
\end{equation}

\subsection{Wideband MU-MIMO Channel}

After SIM$_2$, the transmitted waveform is a standard CP-extended OFDM signal in the physical time domain. The physical subcarrier frequencies are $f_k = f_c + \left(k-\frac{K-1}{2}\right)\Delta f,~k\in\mathcal{K}$, where $\Delta f = B/K$ is the subcarrier spacing over bandwidth $B$. For user $u\in\mathcal{U}\triangleq\{1,\ldots,U\}$, let $P_u$ denote the number of propagation paths, the equivalent frequency-domain channel vector on physical subcarrier $k$ is modeled as
\begin{equation}
    \mathbf{h}_{u}[k]
    =
    \sum_{p=1}^{P_u}
    g_{u,p}(f_k)
    \,
    \mathbf{a}_{t, u, p}^{H}(f_k)
    \in\mathbb{C}^{1\times N},
    \label{eq:hu}
\end{equation}
where $g_{u,p}(f_k)$ is the complex gain of path $p$ and $\mathbf{a}_{t,u,p}(f_k)$ is the transmit steering vector. Collecting all users yields
\begin{equation}
    \mathbf{H}[k]
    \triangleq
    \begin{bmatrix}
        \mathbf{h}_{1}^{\top}[k],
        \mathbf{h}_{2}^{\top}[k],
        \cdots,
        \mathbf{h}_{U}^{\top}[k]
    \end{bmatrix}^{\top}
    \in\mathbb{C}^{U\times N}.
    \label{eq:Hk}
\end{equation}

After propagation through the channel and receiver-side CP removal and DFT at the $U$ users, the received signal on physical subcarrier $k$ is denoted by
\begin{equation}
\mathbf{y}[k]
=
\mathbf{H}[k]\mathbf{P}[k]\mathbf{s}[k]+\mathbf{n}[k],
\label{eq:yk}
\end{equation}
where $\mathbf{n}[k]\sim\mathcal{CN}(\mathbf{0},\sigma^2\mathbf{I})$ and $\sigma^2$ is the noise variance.

It is worth noting that \eqref{eq:yk} represents the nominal equivalent channel assuming an ideal IDFT-and-CP operation. In practice, the finite-resolution quantization of SIM$_2$ inevitably introduces residual inter-carrier interference (ICI) and multi-user interference (MUI). The impact of this residual leakage will be explicitly quantified and evaluated in Section \ref{sec:results}.

\section{Problem Formulation and Solution}

In this section, we use QAT-GD to optimize two coupled operator-fitting problems of SIM$_1$ and SIM$_2$, respectively.

\subsection{SIM$_1$ Fitting to a Block-Diagonal Wideband Precoder}

We set $U\!=\!S\!<\!N$ for a one-stream-per-user transmission model. The desired SIM$_1$ action $\mathbf{F}_{\mathrm{ref,all}}\in\mathbb{C}^{NK\times SK}$ is specified by the block-diagonal wideband precoding target:
\begin{equation}
    \mathbf{F}_{\mathrm{ref,all}}
    \triangleq
    \operatorname{blkdiag}
    \big(
        \mathbf{F}_{\mathrm{ref}}[0],\mathbf{F}_{\mathrm{ref}}[1],\ldots,\mathbf{F}_{\mathrm{ref}}[K-1]
    \big),
    \label{eq:Frefall_bp}
\end{equation}
where $\mathbf{F}_{\mathrm{ref}}[k]=
    \mathbf{H}^{H}[k]
    \big(
        \mathbf{H}[k]\mathbf{H}^{H}[k]
    \big)^{-1}\in\mathbb{C}^{N\times S}$ denotes the reference zero-forcing (ZF) precoder associated with the $k$-th virtual subcarrier index. 

To improve the fitting flexibility across the virtual-subcarrier family, we introduce a block-dependent auxiliary scaling vector $\boldsymbol{\alpha}=[\alpha_0,\alpha_1,\ldots,\alpha_{K-1}]^T\in\mathbb{C}^{K\times1}$ for $K$ blocks in the diagonal precoder. Accordingly, we adopt the following block-wise surrogate objective:
\begin{subequations}
\label{eq:P1_f}
\begin{align}
(\mathcal{P}_1)\quad
\min_{\mathbf{\Phi},\boldsymbol{\alpha}}\;&
J_1
=\sum_{k=0}^{K-1}\left\|\alpha_k\mathbf{P}[k]-\mathbf{F}_{\mathrm{ref}}[k]\right\|_F^2
\label{eq:P1a_new}
\\
\text{s.t.}\;&
e^{j\theta_m^l}\in\mathcal{C}_{\mathrm{SIM1}},\quad \forall l,m ,
\label{eq:P1b_new}
\end{align}
\end{subequations}
where $\mathcal{C}_{\mathrm{SIM1}}$ denotes the feasible set of discrete phases.

\subsection{SIM$_2$ Fitting to the Ideal OFDM Modulation Operator}

Let $\mathbf{F}_{\mathrm{IDFT}}\in\mathbb{C}^{K\times K}$ denote the normalized IDFT matrix with entries
\begin{equation}
[\mathbf{F}_{\mathrm{IDFT}}]_{a,b}
=
\frac{1}{\sqrt{K}}e^{j2\pi ab/K},
\quad a,b\in\mathcal{K},
\end{equation}
and define the CP insertion matrix with CP length $N_{\mathrm{CP}}$
\begin{equation}
\mathbf{C}
=
\begin{bmatrix}
\mathbf{0}_{N_{\mathrm{CP}}\times (K-N_{\mathrm{CP}})} & \mathbf{I}_{N_{\mathrm{CP}}}\\
\mathbf{I}_{K}
\end{bmatrix}
\in\{0,1\}^{T\times K}.
\end{equation}

The ideal CP-extended OFDM synthesis operator for one transmit port is then
\begin{equation}
\mathbf{F}_{\mathrm{cp}}
=
\mathbf{C}\mathbf{F}_{\mathrm{IDFT}}
\in\mathbb{C}^{T\times K}.
\end{equation}

Applying the same OFDM modulation operator to all $N$ transmit dimensions yields
\begin{equation}
\mathbf{Q}_{\mathrm{ideal}}
=
\mathbf{F}_{\mathrm{cp}}\otimes \mathbf{I}_N
\in\mathbb{C}^{NT\times NK},
\label{eq:Qideal_new}
\end{equation}

Hence, no additional permutation matrix is required. Writing \eqref{eq:Qideal_new} in row-block form gives
\begin{equation}
\mathbf{Q}_{\mathrm{ideal}}
=
\big[
\mathbf{Q}_{\mathrm{ideal}}[0]^T,\ldots,\mathbf{Q}_{\mathrm{ideal}}[T-1]^T
\big]^T,
\label{eq:Qideal_blocks}
\end{equation}
where each row block $\mathbf{Q}_{\mathrm{ideal}}[t]\in\mathbb{C}^{N\times NK}$ represents the desired $N$-dimensional port-time-domain output at slot $t$ of one CP-extended OFDM block, rather than a separate spatial precoding matrix.

To capture row-wise fitting mismatch across the $N$ output ports at time slot $t$, we introduce
\begin{equation}
\mathbf{D}_t(\boldsymbol{\beta}_t)
=
\operatorname{diag}(\beta_{t,1},\ldots,\beta_{t,N})
\in\mathbb{C}^{N\times N},
\label{eq:Dbeta_final}
\end{equation}
where $\boldsymbol{\beta}_t=[\beta_{t,1},\ldots,\beta_{t,N}]^T\in\mathbb{C}^{N\times 1}$. The SIM$_2$ fitting problem is then formulated as
\begin{subequations}
\label{eq:P2_f}
\begin{align}
(\mathcal{P}_2)~
\min_{\bar{\mathbf{\Psi}},\{\boldsymbol{\beta}_t\}_{t=0}^{T-1}}\;&
J_2=\sum_{t=0}^{T-1}
\left\|
\mathbf{D}_t(\boldsymbol{\beta}_t)\mathbf{Q}[t]-\mathbf{Q}_{\mathrm{ideal}}[t]
\right\|_F^2
\label{eq:P2_f_a}
\\
\text{s.t.}\;&
e^{j\zeta_\mu^\ell[t]}\in\mathcal{C}_{\mathrm{SIM2}},
\qquad \forall \ell,\mu,t ,  \label{st.psi}
\end{align}
\end{subequations}
where $\mathcal{C}_{\mathrm{SIM2}}$ denotes the feasible set of discrete phase values.

\subsection{Discrete Phase Optimization via QAT-GD}

Problems $\mathcal{P}_1$ and $\mathcal{P}_2$ are nonconvex due to the cascaded multiplicative SIM structure and the discrete phase sets. Direct projected gradient descent (PGD) has some drawbacks here because hard nearest-codeword projection wipes out small gradient steps and often leads to quantization deadlock. We therefore adopt a QAT-GD framework for optimization. 

For a fixed $\mathbf{\Phi}$, $J_1$ in \eqref{eq:P1a_new} decouples across $k$. Setting the derivative to zero, each $\alpha_k$ admits the closed-form update
\begin{equation}
\alpha_k^\star
=
\frac{
\operatorname{tr}\!\big(\mathbf{P}^H[k]\mathbf{F}_{\mathrm{ref}}[k]\big)
}{
\|\mathbf{P}[k]\|_F^2
},
\quad \forall k\in\mathcal{K}.
\label{eq:alpha_closed_final}
\end{equation}

Similarly, for a fixed $\bar{\mathbf{\Psi}}$, $J_2$ in \eqref{eq:P2_f_a} is separable across both the time index $t$ and the output-space index. Hence, the optimal row-wise compensation vector $\boldsymbol{\beta}_t$ satisfies
\begin{equation}
\boldsymbol{\beta}_t^\star
=
\operatorname{diag}\!\big(
\mathbf{Q}_{\mathrm{ideal}}[t]\mathbf{Q}^H[t]
\big)
\oslash
\operatorname{diag}\!\big(
\mathbf{Q}[t]\mathbf{Q}^H[t]
\big),
\quad \forall t\in\mathcal{T},
\label{eq:beta_closed_final}
\end{equation}
where $\operatorname{diag}(\cdot)$ extracts the diagonal entries of the inner matrix into a vector and $\oslash$ denotes the element-wise division.

To update the SIM phases, we maintain hidden continuous variables $\bar{\theta}_m^l$ and $\bar{\zeta}_\mu^\ell[t]$, while the physical discrete phases used in the forward propagation are obtained through element-wise quantization operator $\mathcal{Q}(\cdot)$:
\begin{equation}
\theta_m^l=\mathcal{Q}_{\mathrm{SIM1}}(\bar{\theta}_m^l),
\qquad
\zeta_\mu^\ell[t]=\mathcal{Q}_{\mathrm{SIM2}}(\bar{\zeta}_\mu^\ell[t]).
\label{eq:quant}
\end{equation}

To update the phase shifts of the $l$-th SIM$_1$ layer, we fix all other layers and rewrite the operator as 
\begin{equation}
\mathbf{P}[k]
=
\mathbf{A}_{l,k}\mathbf{\Phi}^{l}\mathbf{B}_{l,k},
\label{eq:P_factor_final}
\end{equation}
where the left item $\mathbf{A}_{l,k}\in\mathbb{C}^{N\times NK}$ and the right item $\mathbf{B}_{l,k}\in\mathbb{C}^{NK\times S}$ are the corresponding block-restricted factors. Defining the block-wise residual $\mathbf{E}_k = \alpha_k\mathbf{P}[k]-\mathbf{F}_{\mathrm{ref}}[k]$, the layer-wise gradient of \eqref{eq:P1a_new} used in QAT-GD is given by
\begin{equation}
\nabla_{\boldsymbol{\theta}^{l}}J_1
=
-2\,\operatorname{Im}
\left\{
\boldsymbol{\phi}^{l}\odot
\operatorname{diag}(\mathbf{G}_{l}^{(1)})
\right\},
\label{eq:grad_sim1_final}
\end{equation}
where $\boldsymbol{\phi}^{l} = [e^{j\theta_1^l},\ldots,e^{j\theta_{NK}^l}]^T$ is the physical phase vector of SIM$_1$'s $l$-th layer, and $\mathbf{G}_{l}^{(1)} = \sum_{k=0}^{K-1} \alpha_k\,\mathbf{B}_{l,k}\mathbf{E}_k^H\mathbf{A}_{l,k}$.

Similarly, for fixed time slot $t$ and layer $\ell$ in SIM$_2$, we write
\begin{equation}
\mathbf{Q}[t] = \widetilde{\mathbf{A}}_{\ell,t}\mathbf{\Psi}^{\ell}[t]\widetilde{\mathbf{B}}_{\ell,t}.
\label{eq:Q_factor_final}
\end{equation}
where $\tilde{\mathbf A}_{\ell,t}$ and $\tilde{\mathbf B}_{\ell,t}$ denote the left and right cascade factors obtained by fixing all SIM$_2$ layers except $\mathbf{\Psi}^{\ell}[t]$.

Then, we define the residual $\mathbf{E}_t = \mathbf{D}_t(\boldsymbol{\beta}_t)\mathbf{Q}[t] - \mathbf{Q}_{\mathrm{ideal}}[t]$. The surrogate layer-wise gradient of $J_2$ with respect to the phase vector $\boldsymbol{\psi}^{\ell}[t]=[e^{j\zeta_1^\ell[t]},\ldots,e^{j\zeta_{NK}^\ell[t]}]^T$ is given by
\begin{equation}
\nabla_{\boldsymbol{\zeta}^{\ell}[t]}J_2 = -2\,\operatorname{Im} \left\{ \boldsymbol{\psi}^{\ell}[t]\odot \operatorname{diag}(\mathbf{G}_{\ell,t}^{(2)}) \right\},
\label{eq:grad_sim2_final}
\end{equation}
where $\mathbf{G}_{\ell,t}^{(2)} = \widetilde{\mathbf{B}}_{\ell,t}\mathbf{E}_t^H \mathbf{D}_t(\boldsymbol{\beta}_t) \widetilde{\mathbf{A}}_{\ell,t}$.

During the backward pass, the straight-through estimator bypasses the non-differentiable quantizers in \eqref{eq:quant} and updates the hidden continuous variables as
\begin{align}
\bar{\boldsymbol{\theta}}^{\ell,(r+1)}
=
\bar{\boldsymbol{\theta}}^{\ell,(r)}
-
\mu_\ell^{(r)}\nabla_{\boldsymbol{\theta}^{\ell}}J_1^{(r)},
\label{eq:theta_update} \\
\bar{\boldsymbol{\zeta}}^{\ell,(r+1)}[t]
=
\bar{\boldsymbol{\zeta}}^{\ell,(r)}[t]
-
\eta_{\ell,t}^{(r)}\nabla_{\boldsymbol{\zeta}^{\ell}[t]}J_2^{(r)},
\label{eq:zeta_update}
\end{align}
where $\mu_\ell^{(r)}$ and $\eta_{\ell,t}^{(r)}$ are adaptive step sizes. 

In each outer iteration, the auxiliary variables $\{\alpha_k\}$ and $\{\beta_t\}$ are first updated in closed form, and then the hidden continuous phase variables are refined by several QAT-GD steps. The outer loop is terminated when the relative decrease of the objective becomes smaller than a prescribed threshold or when the maximum number of iterations is reached.

\section{Results and Analysis}
\label{sec:results}

We evaluate the proposed SIM-enabled transmitter from three complementary perspectives. First, we examine whether the proposed QAT-GD effectively addresses the optimization difficulty introduced by discrete phase constraints. Second, we verify whether the fitted SIM$_1$ and SIM$_2$ accurately realize the wideband fully-analog precoding and OFDM modulation functionalities. Third, we analyze the resulting MU-MIMO OFDM performance and quantify the performance gap relative to the ideal digital benchmark. In our simulations, we set $N\!=\!6$, $S\!=\!4$, $K\!=\!16$, $N_{\mathrm{CP}}\!=\!4$, $L_1\!=\!7$, and $L_2\!=\!11$.

\subsection{Optimization Behavior under Discrete Phase Constraints}

\begin{figure}[t]
    \centering
    \subfigure[SIM$_1$ convergence performance.\label{fig:subfig1a}]{\includegraphics[width=0.253\textwidth]{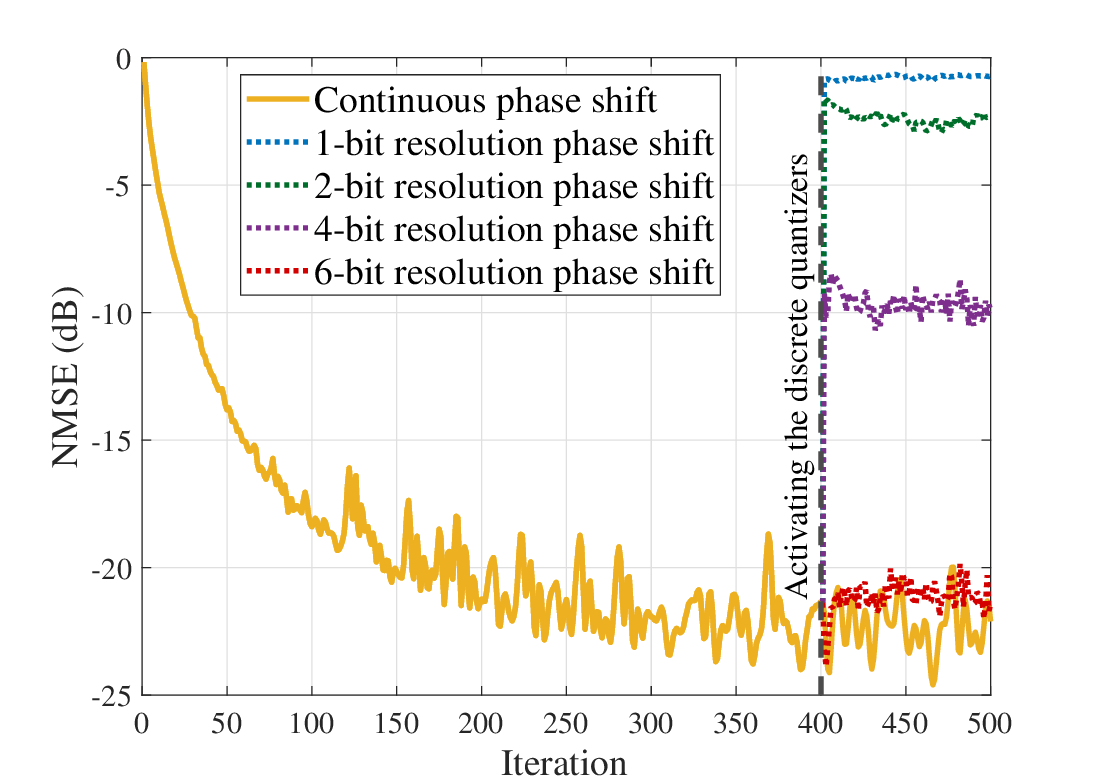}}
    \hspace{-0.6 cm}
    \subfigure[SIM$_1$ heatmap comparison.\label{fig:subfig1b}]{\includegraphics[width=0.253\textwidth]{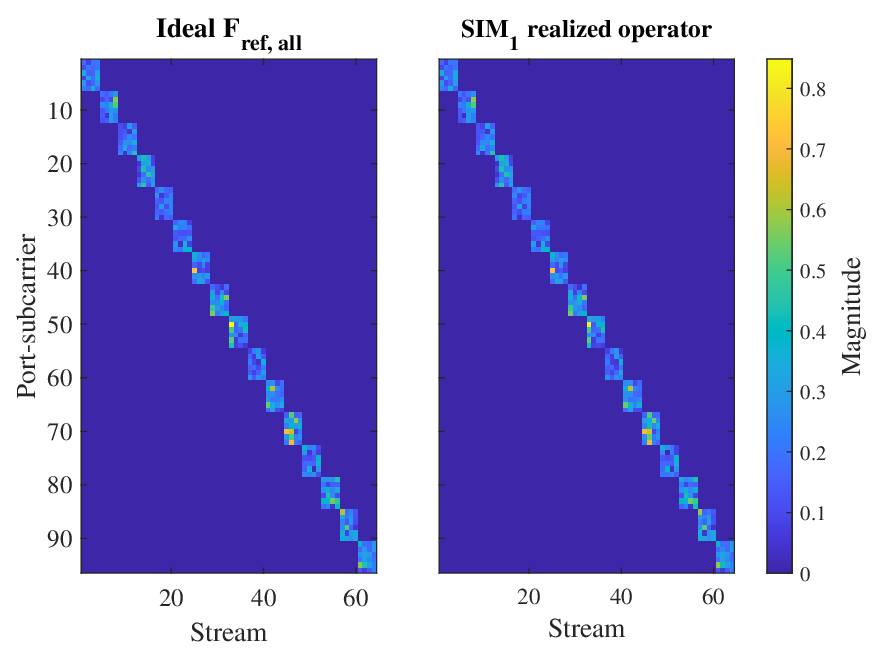}}
    \subfigure[SIM$_2$ convergence performance.\label{fig:subfig1c}]{\includegraphics[width=0.253\textwidth]{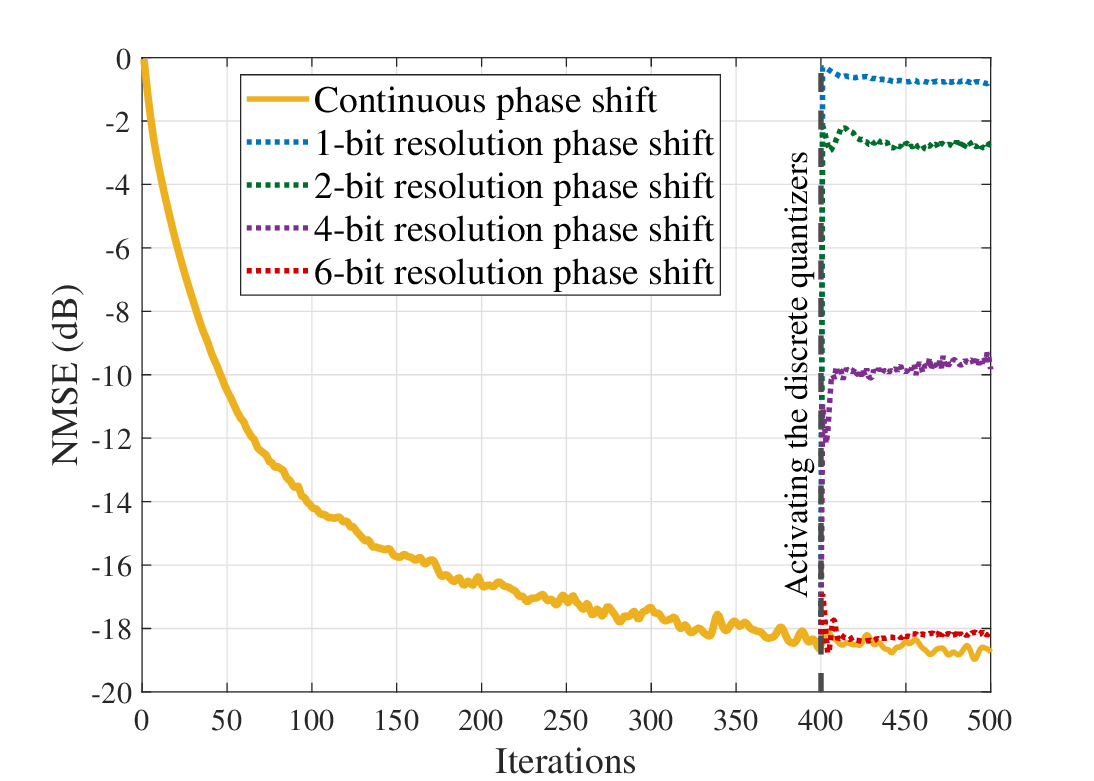}}
    \hspace{-0.6 cm}
    \subfigure[SIM$_2$ heatmap comparison.\label{fig:subfig1d}]{\includegraphics[width=0.253\textwidth]{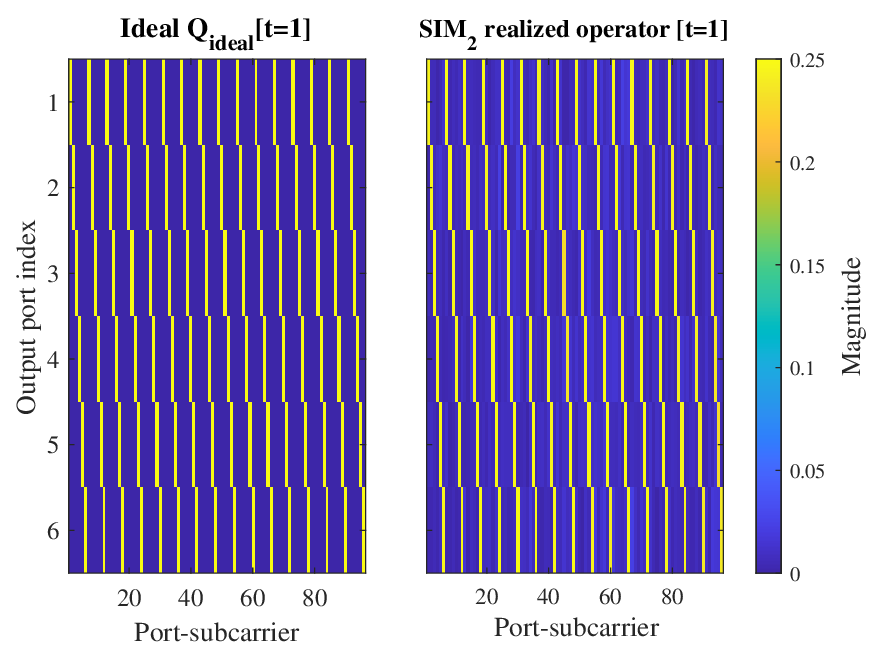}}
\caption{{Architecture validity of the cascaded wave-domain transmitter under various discrete phase constraints. }}
    \label{fig:operator_fitting}
\vspace{-0.5 cm}
\end{figure}

Fig. \ref{fig:subfig1a} and Fig. \ref{fig:subfig1c} depict the normalized mean square error (NMSE) convergence behaviors for SIM$_1$ and SIM$_2$, respectively. A continuous-phase warm-up stage is first used to drive the solution toward a favorable basin during offline, after which the discrete phase quantizers are activated and the optimization continues under the target quantized constraint. Low-resolution phase shifters, such as 1-bit and 2-bit, suffer noticeable quantization deadlock and therefore exhibit visible NMSE floors, whereas higher-resolution settings move much closer to the continuous-phase lower bounds. In particular, 6-bit quantization enables both cascades to converge stably to low-error solutions, demonstrating the effectiveness of the proposed QAT-GD under practical discrete phase constraints. This is because our proposed baseband-free wideband architecture imposes substantially more stringent orthogonality constraints, which need high-resolution phase shifts for close realization.

The heatmap comparisons in Fig. \ref{fig:subfig1b} and Fig. \ref{fig:subfig1d} further confirm this conclusion. For SIM$_1$, the realized operator closely reproduces the desired block-diagonal wideband precoding structure, while the residual off-block leakage remains weak. For SIM$_2$, the realized operator accurately captures the sparse comb-like structure of the target IDFT-and-CP modulation operator at the representative time slot. The staggered periodic bright spots reflect the desired per-port subcarrier selectivity with only minor residual leakage. Together, these results indicate that both wideband MU-MIMO precoding and OFDM modulation can be closely approximated in the EM wave domain using 6-bit discrete phase shifts.

\subsection{Precoder and OFDM Functionality Validation}

\begin{figure}[t]
    \centering
    \subfigure[The 3D magnitude plot.\label{fig:subfig2a}]{\includegraphics[width=0.24\textwidth]{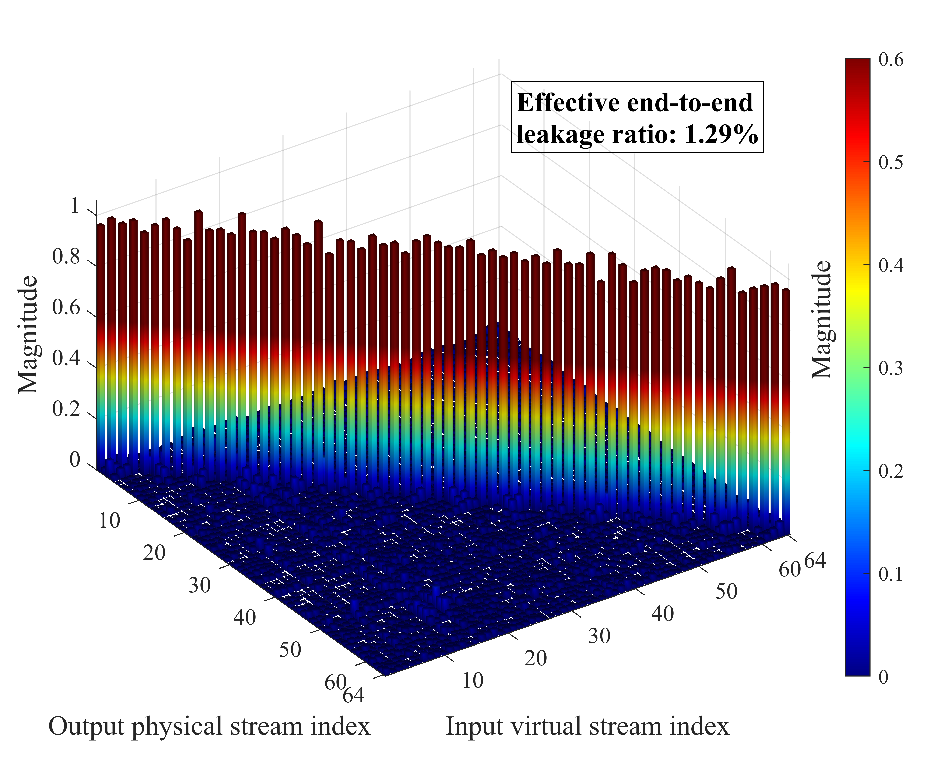}}
    \hspace{-0.3 cm}
    \subfigure[The 2D diagnostic heatmap.\label{fig:subfig2b}]{\includegraphics[width=0.24\textwidth]{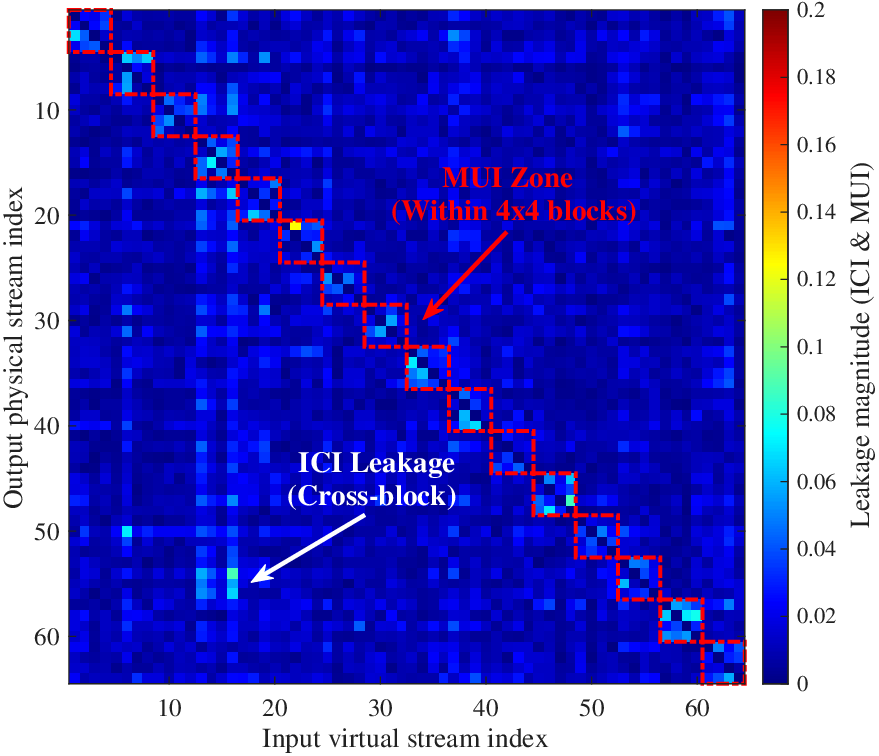}}
\caption{{Validation of the wave-domain OFDM modulation, where the off-diagonal spatial dispersion vividly distinguishes the localized MUI from the global ICI.}}
    \label{fig:ofdm_validation}
\vspace{-0.4 cm}
\end{figure}

Fig.~\ref{fig:ofdm_validation} validates the functional correctness of the proposed wave-domain transmitter. As depicted in the 3D magnitude plot of Fig. \ref{fig:subfig2a}, the effective operator exhibits a towering diagonal ridge accompanied by an exceptionally low off-diagonal noise floor. To better interpret the residual leakage, Fig.~\ref{fig:subfig2b} maps the energy distribution of the nominal 1.29\% aggregate leakage by intentionally suppressing the dominant main diagonal to zero. The residual interference can be structurally decoupled into two distinct physical error sources: intra-block MUI, caused by imperfect SIM$_1$ spatial decoupling, and inter-block ICI, caused by imperfect SIM$_2$ OFDM modulation. As visually corroborated, the sparse off-block ICI is effectively suppressed, preserving nearly 98.71\% of the signal energy on the desired orthogonal dimensions. Despite the severe physical constraints of passive wave propagation and 6-bit discrete quantization, this indicates that the proposed architecture can closely realize the intended OFDM modulation functionality in the EM wave domain while largely preserving subcarrier orthogonality.

\subsection{MU-MIMO OFDM Performance Analysis}

\begin{figure}
\centerline{\includegraphics[width=0.45\textwidth]{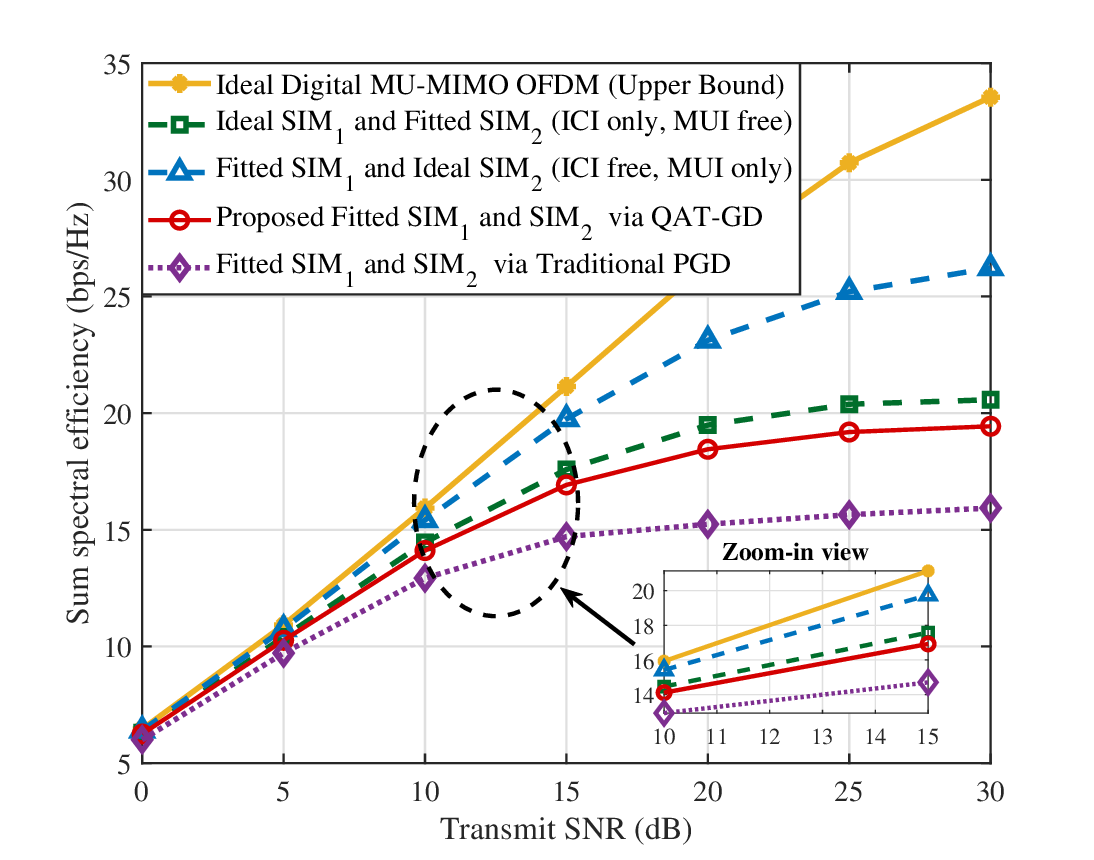}}
\caption{Sum spectral efficiency~versus~SNR, where the performance gap is decomposed into the spatial precoding loss (SIM$_1$) and the temporal OFDM modulation loss (SIM$_2$). }
\label{fig:sumrate}
\vspace{-0.4 cm}
\end{figure}

Fig. \ref{fig:sumrate} evaluates the sum spectral efficiency versus the transmit SNR. In addition to the ideal digital MU-MIMO OFDM upper bound, we include two ablation baselines:~i)~cascade of ideal SIM$_1$ and fitted SIM$_2$ and ii) cascade of fitted SIM$_1$ and ideal SIM$_2$, in order to examine the losses introduced by the two wave-domain stages. We also compare against a traditional PGD baseline under the same discrete phase constraint.

In the low-to-medium SNR regime (e.g., $\le 15$ dB), the proposed full SIM cascade closely tracks the ideal digital benchmark, indicating that the residual wave-domain leakage is sufficiently small when the system is noise-limited. As the SNR increases, the system gradually becomes interference-limited, and the gap to the ideal digital benchmark becomes more visible. The ablation results show that the temporal OFDM modulation loss induced by SIM$_2$ is more critical than the spatial precoding loss induced by SIM$_1$, highlighting the higher sensitivity of OFDM waveform synthesis to discrete phase quantization. Nevertheless, compared with direct projected-gradient optimization, the proposed QAT-GD substantially improves the achievable sum rate, confirming its practical necessity for discrete-phase wave-domain transmitter design. Overall, these results support the feasibility of physically realizing structured transmitter-side baseband functions through programmable wave-domain computation.

\section{Conclusions}
\label{sec:conclusion}
This paper proposed a novel baseband-free wideband MU-MIMO OFDM transmitter architecture where symbol loading, fully-analog precoding, and OFDM modulation are redistributed across two cascaded SIMs. By introducing a virtual-subcarrier coefficient representation under monochromatic excitation, SIM$_1$ performs channel-adaptive stream-to-port mapping with a block-diagonal wideband precoding target, while SIM$_2$ generates the CP-extended OFDM waveform through the IDFT-and-CP operator. Under practical discrete phase constraints, the proposed QAT-GD framework effectively enables operator fitting for both metasurface stages. Overall, the proposed architecture shows that wideband MU-MIMO OFDM transmission can be reinterpreted as a wave-domain computing problem, thereby opening a new path toward architecturally compact and baseband-free wireless transmitters.

\end{document}